\documentclass{article}
\usepackage{amsmath}

\begin{document}

\begin{center}
The Systematic Measurement Errors and Uncertainty Relation

Timur F. Kamalov

Physics Department, Moscow State Open University

Korchagina, 22, Moscow, 107996, Russia

E-mail: TimKamalov@mail.ru
\end{center}

\textit{Inertial effects in non-inertial reference frames are compared with
quantum properties of tests objects. The real space-time and perfect
inertial reference frame can be compared accurate to the uncertainty
relation. Complexities if describing micro-object in non-inertial
reference-frames are avoidable using Ostrogradski's Canonical Formalism.}



Newtonian mechanics is the simplest version of describing mechanical systems
in inertial reference frames when the effect of higher derivatives of
coordinates with respect time is negligible. However, perfect inertial
systems are very difficult to obtain in practice. A simplified consideration
of the real reference frame as inertial, permits one to obtain equations of
motion usually solved by traditional methods of mathematical physics.
However, in the general case, while considering some real reference frame,
to describe dynamics of body motion in a more exist approximation,
consideration of complex problems with the equations not easily solved.
Neglecting the effect of inertial forces on test bodies in a non-inertial
reference frame, a systematic error of measuring the coordinate and momentum
cannot be obtain.

To describe dynamics of body motion in any reference frames, we expand the
function $r=r(t)$ in a Taylor series

\begin{center}
\begin{equation}
r=r_{0}+vt+\frac{at^{2}}{2}+\frac{1}{3!}\overset{\cdot }{a}t^{3}+\frac{1}{4!}%
\overset{\cdot \cdot }{a}t^{4}+...+\frac{1}{n!}\overset{\cdot }{r}%
^{(n)}t^{n}+...  \label{eq:1}
\end{equation}
\end{center}

denoting the position on space by $r$, and time by $t$.

The kinematic formula of classical physics in inertial reference frames
restricts consideration to only second derivatives of $r$ with respect the
time (acceleration $a)$%
\begin{equation}
r_{Newton}=r_{0}+vt+\frac{at^{2}}{2}.  \label{eq:2}
\end{equation}

Denoting the hidden variables accounting for additional terms in (1) with
respect to (2) as $q_{r}$,
\begin{equation}
q_{r}=\frac{1}{3!}\overset{\cdot }{a}t^{3}+\frac{1}{4!}\overset{\cdot \cdot }%
{a}t^{4}+...+\frac{1}{n!}\overset{\cdot }{r}^{(n)}t^{n}+...  \label{eq:3}
\end{equation}

rewrite (1) in the form%
\begin{equation}
r=r_{Newton}+q_{r}
\end{equation}

Comparing the action function for cases (1) and (2)%
\begin{equation}
S-S_{Newton}=nh
\end{equation}

we denote the upper bound of the difference action functions as $h$. Here $S$
is the action function in any real reference frames (including inertial) and
$S_{Newton}$ - in inertial reference frames.

Then the uncertainty due incompleteness of the description particles in a
real reference frames and in inertial reference systems:%
\begin{equation}
mr\overset{\cdot }{r}-mr_{Newton}\overset{\cdot }{r}%
_{Newton}=m(r-r_{Newton})(\overset{\cdot }{r}-\overset{\cdot }{r}%
_{Newton})=nh,
\end{equation}

where $h$ being an upper bound of hidden action in inertial reference
frames. Higher time derivatives of spatial coordinates act as hidden and
addition variables complementing the description of sample particles for
inertial reference systems.

Ostrogradski's avoided the above disadvantages of Newtonian mechanics,
having assumed the Lagrange function to depend on not only first derivatives
of coordinates with respect to time, but also on higher derivatives. There
are no inertial reference frames in Ostrogradski's model, and it can
describe complex motion with higher derivatives of accelerations. Such a
model is known in the literature as Ostrogradski's Canonical Formalism.

For inertial reference systems the Lagrangian $L$ is the function of only
the coordinates and their first derivatives, $L=L(t,r,\overset{\cdot }{r})$
For non-inertial reference systems, the Lagrangian depends on the
coordinates and their higher derivatives as well as of the first one, i.e. $%
L=L(t,r,\overset{\cdot }{r},\overset{\cdot \cdot }{r},\overset{\cdot \cdot
\cdot }{r},...,\overset{\cdot (n)}{r})$ Applying the principle of least
action, we get [2]%
\begin{equation}
\delta S=\delta \int L(r,\overset{\cdot }{r},\overset{\cdot \cdot }{r},%
\overset{\cdot \cdot \cdot }{r},...,\overset{\cdot (n)}{r})dt=\int
\sum_{n=0}^{N}(-1)^{n}\frac{d^{n}}{dt^{n}}(\frac{\partial L}{\partial
\overset{\cdot (n)}{r}})\delta rdt=0\text{.}
\end{equation}%
Then, the Euler -- Lagrange function for complex non-inertial reference
systems takes on the form

\begin{center}
\begin{equation}
\frac{\partial L}{dr}-\frac{d}{dt}(\frac{\partial L}{\partial \overset{\cdot
}{r}})+\frac{d^{2}}{dt^{2}}(\frac{\partial L}{\partial \overset{\cdot \cdot }%
{r}})-\frac{d^{3}}{dt^{3}}(\frac{\partial L}{\partial \overset{\cdot \cdot
\cdot }{r}})+\frac{d^{4}}{dt^{4}}(\frac{\partial L}{\partial \overset{\cdot
(4)}{r}})+...=0
\end{equation}
\end{center}

Or

\begin{center}
\begin{equation}
\sum_{n=0}^{N}(-1)^{n+1}\frac{d^{n}}{dt^{n}}(\frac{\partial L}{\partial
\overset{\cdot (n)}{r}})=0
\end{equation}
\end{center}

Denoting

\begin{center}
$F=\frac{\partial L}{\partial r},p=\frac{\partial L}{\partial \overset{\cdot
}{r}}$

$F^{(1)}=\frac{\partial \overset{\cdot \cdot }{L}}{\partial \overset{\cdot
\cdot }{r}},p^{(1)}=\frac{\partial L}{\partial \overset{\cdot \cdot \cdot }{r%
}}$

$F^{(2)}=\frac{\partial \overset{\cdot (4)}{L}}{\partial \overset{\cdot (4)}{%
r}},p^{(2)}=\frac{\partial L}{\partial \overset{\cdot (5)}{r}}$

..................

$F^{(\alpha )}=\frac{\partial \overset{\cdot (2\alpha )}{L}}{\partial
\overset{\cdot (2\alpha )}{r}},p^{(\alpha )}=\frac{\partial L}{\partial
\overset{\cdot (2\alpha +1)}{r}}$.
\end{center}

we get the description of inertial forces for complex non-inertial reference
systems. The value of the total force taking into account the Coriolis force
may be expressed through momentums in non-inertial reference systems and
their derivatives:

\begin{center}
\begin{equation}
F+F^{(1)}+F^{(2)}+...+F^{(\alpha )}=\frac{dp}{dt}+\frac{d^{2}p^{(1)}}{dt^{2}}%
+\frac{d^{3}p^{(2)}}{dt^{3}}+...+\frac{d^{\alpha +1}p^{(\alpha )}}{%
dt^{\alpha +1}}
\end{equation}
\end{center}

Denoting the energy brought about by the non-inertial reference system as $Q$
and the constant coefficients as $\alpha _{i}$, we get for the potential
energy $V$ and kinetic energy $W$ the following expressions:

\begin{center}
$E=V+W+Q$

$V=\alpha _{1}r^{2}$

$W=\alpha _{2}\overset{\cdot }{r}^{2}$

$Q=\alpha _{2}\overset{\cdot \cdot }{r}^{2}+...+\alpha _{n}\overset{\cdot (n)%
}{r}^{2}+...$
\end{center}

Jacobi-Hamilton equation for the action function takes the form:

\begin{center}
\begin{equation}
-\frac{\partial S}{\partial t}=\frac{(\nabla S)^{2}}{2m}+V+Q,
\end{equation}
\end{center}

and let us call $Q$ the quantum potential. Here, is the velocity $v=\frac{%
\partial S}{\partial t}=\frac{\nabla S}{m}$, and the acceleration $a=\overset%
{\cdot }{v}=\frac{\nabla \overset{\cdot }{S}}{m}=\frac{\nabla ^{2}S}{m}$,
where is the continuity equation $\frac{\partial v}{\partial t}+v\frac{%
\partial v}{\partial r}=0$ for the vector $v$. Here is $\overset{\cdot \cdot
}{v}=\nabla \overset{\cdot \cdot }{S}$. Then, denoting $\frac{\partial S}{%
\partial \overset{\cdot (n)}{r}}=p^{(n-1)}$, we get the following equation
for non-inertial reference systems:

\begin{center}
$\frac{dS}{dt}=\frac{\partial S}{\partial t}+\frac{\partial S}{\partial r}%
\frac{\partial r}{\partial t}+\frac{\partial S}{\partial \overset{\cdot }{r}}%
\frac{\partial \overset{\cdot }{r}}{\partial t}+...+\frac{\partial S}{%
\partial \overset{\cdot (n)}{r}}\frac{\partial \overset{\cdot (n)}{r}}{%
\partial t}$,

$\frac{dS}{dt}=\frac{\partial S}{\partial t}+\frac{(\nabla S)^{2}}{2m}%
+p^{(0)}a+p^{(1)}\overset{\cdot }{a}+...+p^{(n)}\overset{\cdot (n)}{a}$
\end{center}

Therefore, the equation of motion of a sample particle in a complex
non-inertial reference system shall be as follows:

\begin{center}
\begin{equation}
\frac{dS}{dt}=\frac{\partial S}{\partial t}+\frac{(\nabla S)^{2}}{2m}+Q
\end{equation}
\end{center}

Here is $Q=p^{(0)}a+p^{(1)}\overset{\cdot }{a}+...+p^{(n)}\overset{\cdot (n)}%
{a}$.

In the first approximation $Q\approx \alpha _{3}\frac{\nabla ^{2}S}{m^{2}}.$
(the constant is chosen as $\alpha _{3}=\frac{i\hbar m}{2}$). Hence, we get
the Schrodinger equation in the first approximation for the function $\psi
=e^{\frac{i}{\hbar }S}$.

The distinctive feature of above model lies in an uncertainty of coordinate
and momentum of a microobject being explicable not only by quantum axiomatic
but also by imperfectness of the reference frame due to the effect, e.g.
relict stationary or static gravitational fields for example. This means
that we consider non-inertial reference frames, and the uncertainty in the
position of the particle being examined can be related to fluctuations of
the reference body and the reference frame connected with it. To estimate
such an effect, it is necessary to compare the systematic error of
measurements with the quantum uncertainties of the coordinate and momentum
in the uncertainty relation.

\begin{center}
\textit{References}
\end{center}

[1] Lagrange J.I., Mecanique analitique. Paris, De Saint, 1788.

[2] Ostrogradskii M., Met. De l'Acad. De St.-Peterburg, v. 6, p. 385, 1850.

\end{document}